\documentclass[twocolumn,floatfix,pra,aps,showpacs,superscriptaddress]{revtex4}
\usepackage{epsfig,graphicx}
\usepackage{flafter}
\usepackage{amsmath}
\usepackage{color}
\usepackage{bm}
\usepackage{amssymb}
\usepackage{epstopdf}
\usepackage{amstext}
\usepackage{amsthm}
\usepackage{amsfonts}
\usepackage{latexsym}
\usepackage{multirow}
\usepackage{mathtools}
\usepackage{bm}
\usepackage{bbm}

\begin{document}

\title{Decay of the relative phase domain wall into confined vortex pairs: 
the case of a coherently coupled bosonic mixture}

\author{A. Gallem\'{\i}}
\affiliation{INO-CNR BEC Center and Dipartimento di Fisica, Universit\`a degli Studi di Trento, 38123 Povo, Italy}
\affiliation{Trento Institute for Fundamental Physics and Applications, INFN, 38123, Trento, Italy}
\author{L. P. Pitaevskii}
\affiliation{INO-CNR BEC Center and Dipartimento di Fisica, Universit\`a degli Studi di Trento, 38123 Povo, Italy}
\affiliation{Kapitza Institute for Physical Problems RAS, Kosygina 2, 119334 Moscow, Russia}
\author{S. Stringari}
\affiliation{INO-CNR BEC Center and Dipartimento di Fisica, Universit\`a degli Studi di Trento, 38123 Povo, Italy}
\affiliation{Trento Institute for Fundamental Physics and Applications, INFN, 38123, Trento, Italy}
\author{A. Recati}
\affiliation{INO-CNR BEC Center and Dipartimento di Fisica, Universit\`a degli Studi di Trento, 38123 Povo, Italy}
\affiliation{Trento Institute for Fundamental Physics and Applications, INFN, 38123, Trento, Italy}

\date{\today}

\begin{abstract}
A domain wall of relative phase in a flattened harmonically-trapped Bose-Einstein condensed mixture 
of two atomic hyperfine states, subject to a stationary Rabi coupling of intensity $\Omega$, is 
predicted to decay through two different mechanisms. For small values of $\Omega$ the instability 
has an energetic nature and is associated with the formation of a vortex-antivortex pair of the 
same atomic hyperfine states, whose motion inside the trap causes  the emergence of magnetization, 
the bending of the domain wall and its consequent fragmentation. For large values of $\Omega$ 
the domain wall instead undergoes a dynamic snake instability, caused by the negative value of its 
effective mass and results in the fast fragmentation of the wall into smaller domain walls 
confining vortex pairs of different atomic species. Numerical predictions are given by solving 
the time-dependent Gross-Pitaevskii equation in experimentally available configurations of mixtures 
of sodium atomic gases.
\end{abstract}


\maketitle

The study of topological defects in ordered phases encompasses different fields of physics 
from quantum fluids and superconductors to cosmology. Of particular interest is the case of 
composite defects due to the presence of multiple symmetry breaking \cite{Bunkov2000}. In 
this respect ultra-cold gases mixtures are emerging as one of the most suitable platforms 
for studying the dynamical behaviour of topological defects and in particular their formation 
and decay. The easiest system showing interesting composite defects is most probably a 
binary Bose-Bose mixture with an interconversion term, realized via external coherent Rabi 
coupling, between two hyperfine components \cite{Matthews1999b,Zibold2010,Nicklas2011,Nicklas2015}. 
The intriguing features of such a gas are related to the controlled  -- {\textsl{via}} the 
tunability of the Rabi coupling --  breaking of the conservation of the relative atom number. 
Only the total atom number is conserved, due to the $U(1)$ symmetry related global phase
(see, e.g., \cite{Abad2013} and reference therein).   

An interesting solitonic solution is represented by the sine-Gordon domain wall of the relative 
phase, explored by Son and Stephanov \cite{Son2002} for Bose-Einstein condensed gases and by 
Tanaka \cite{Tanaka2001} for two-band superconductors. The relative phase domain wall (hereafter 
simply called domain wall) is characterized by an asymptotic $2\pi$ jump of the relative phase 
between the two condensates, whose space gradient is localized in a narrow region fixed by the 
strength of the Rabi coupling, which makes it energetically costly. Such a solitonic 
configuration is particularly relevant because it provides non trivial constraints on the 
topology of quantized vortices. Indeed, in the presence of coherent coupling between the two 
condensates, a quantized vortex in one of the two components of the mixture, also known as 
half-quantized vortex, has always a domain wall attached to it. As a consequence isolated 
half-quantized vortices cannot exist, but are rather paired in order to screen the relative 
phase change. This situation is reminiscent of the problem of quarks in particle physics, 
which cannot exist as individual particles, but are confined by strings. The analogy with 
the physics of quarks and strings is further reinforced by the linear increase of the confining 
potential with the distance between the vortices in a pair and the possibility for the domain 
wall to connect either a vortex and an antivortex in the same atomic species or vortices with 
the same angular momentum in different atomic species (such pairs are named mesons and baryons, 
respectively, in the recent work by Eto and Nitta \cite{Eto2018}).

Because of the high energy cost, it is preferable for long domain walls to decay into smaller 
ones connecting vortex pairs \cite{Son2002,Gallemi2016,Tylutki2016,Eto2018}. This is expected 
to be the relevant decay mechanism for small values of the Rabi coupling. If instead the Rabi 
coupling is large, domain walls become dynamically unstable \cite{Son2002} and exhibit snake 
instability. Understanding the mechanism of fragmentation of the domain wall is of crucial 
importance in view of future experiments aimed to investigate the formation and the dynamics 
of confined vortex pairs in ultracold atomic gases.

In this Letter we first  provide a thorough discussion of the equilibrium and dynamical properties 
of the domain wall in uniform matter. Then, by carrying out a fully-numerical analysis based on 
the solution of the time-dependent Gross-Pitaevskii equations (GPEs), we explore the decay 
mechanisms of a domain wall in a mixture trapped by a harmonic potential, suggesting a natural 
protocol for possible experiments. As we show in this Letter, the decay of a domain wall into 
shorter domain walls can be monitored not only by measuring the relative phase of the gas, but 
also through the total density and the magnetization profiles.

\begin{figure}[t!]
\includegraphics[width=\linewidth]{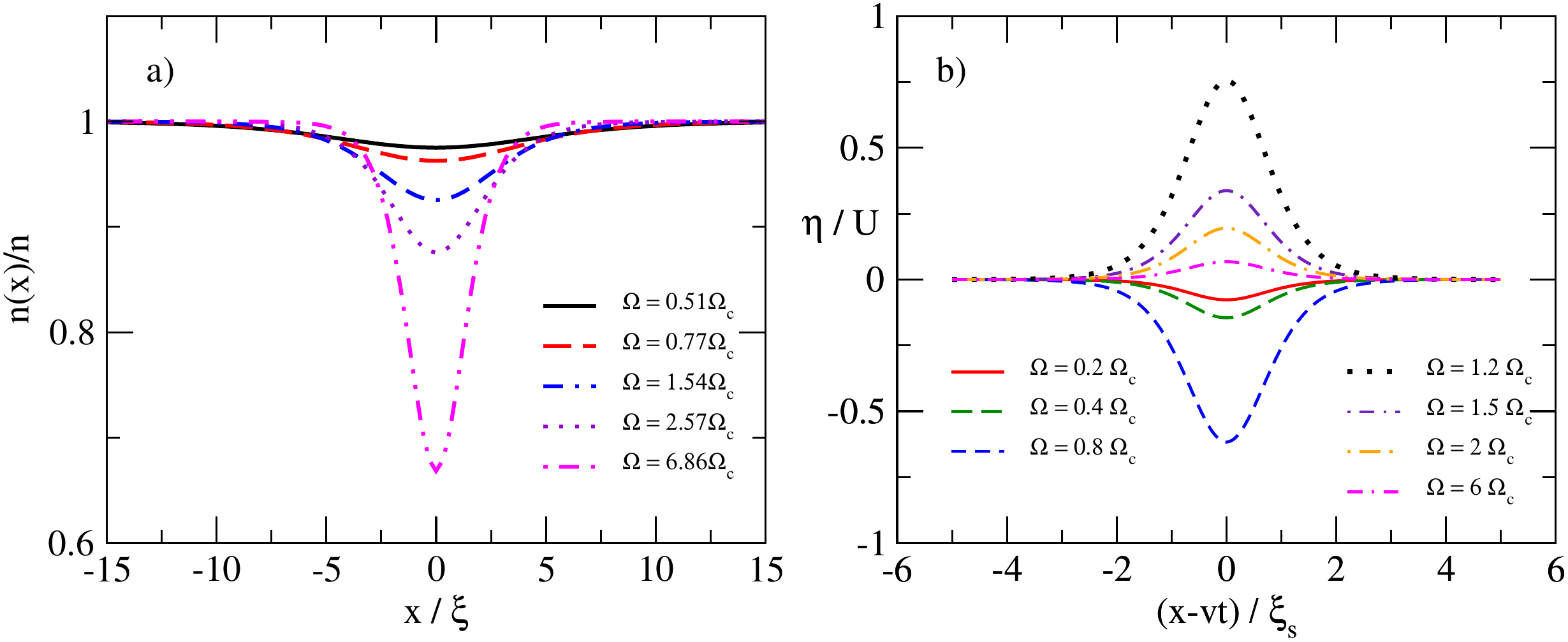}
\includegraphics[width=\linewidth]{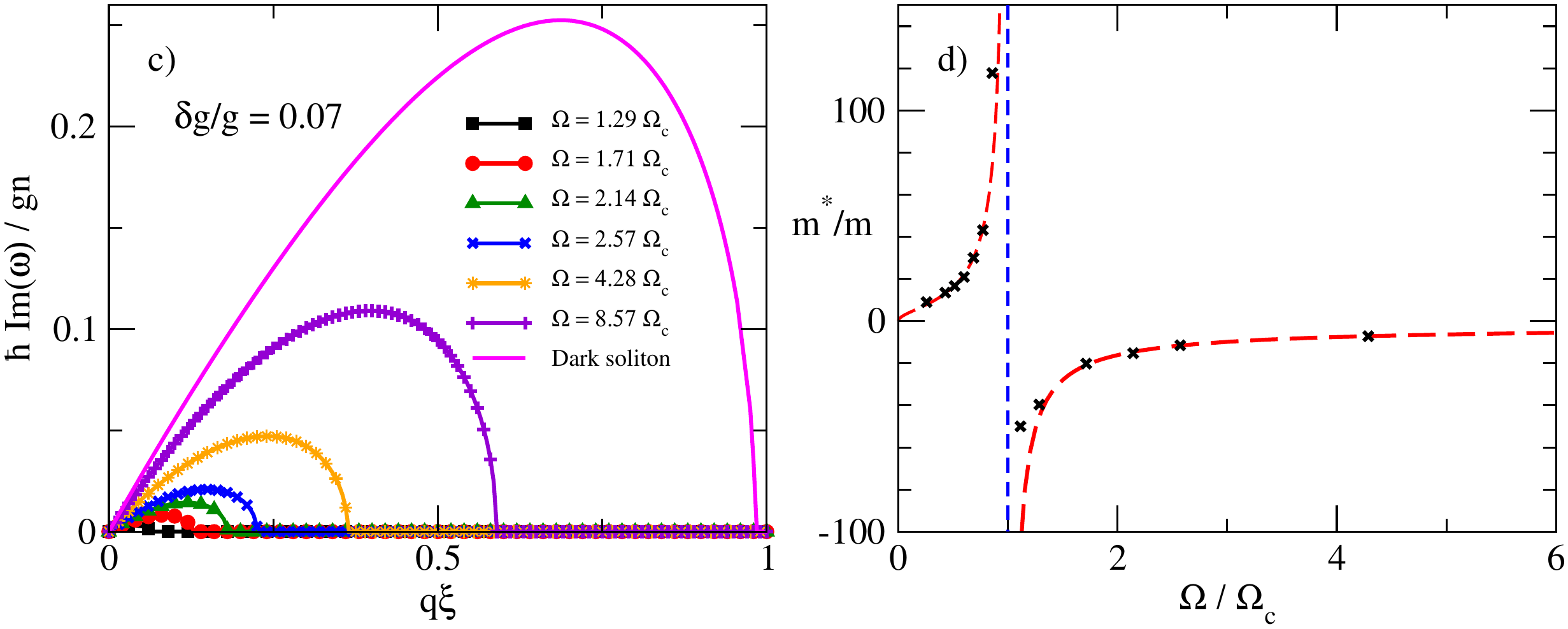}
\caption{a) Plot of the total density as a function of the position of a 1D domain wall for 
different values of $\Omega$. b) The ratio between the magnetization $\eta$ and the 
velocity $U=v/c_s$ as a function of the position, for different values of $\Omega$. c) Plot 
of the imaginary frequencies that appear for $\Omega>\Omega_c$ as a function of the wave vector 
$q$ for different values of $\Omega$. d) Effective mass $m^*$ as a function of $\Omega/\Omega_c$ 
calculated using Eq. (\ref{m*3}) (black crosses), showing a divergence at the critical point 
(vertical blue dashed line). The red dashed line corresponds to the analytical calculation of 
the effective mass of the domain wall in homogeneous matter (\ref{m*2}). In all the panels, 
$\delta g/g=0.07$.}
\label{fig:fig1}
\end{figure}

We consider a zero temperature mixture of two hyperfine states of a bosonic atomic isotope, 
which are coherently-coupled through a Rabi term $\Omega$. Defining  $(\Psi_1,\Psi_2)$ the 
order parameter of the mixture, the grand canonical energy functional takes the form: 
\begin{align} 
\mathcal{E}_{\rm GP}=&\sum_{i=1,2}\left[\frac{\hbar^2}{2m}|\nabla\Psi_i|^2+(V_{\rm ext}-\mu)
|\Psi_i|^2+\frac{g_{ii}}{2}|\Psi_i|^4\right]\nonumber\\
&+g_{12}|\Psi_1|^2|\Psi_2|^2-\frac{1}{2}\hbar\Omega\left(\Psi^*_1\Psi_2+\Psi_2^*\Psi_1\right)\,,
\label{eq:eq1}
\end{align}
where $m$ is the atomic mass, $\mu$ is the chemical potential and $V_{\rm ext}$ an external 
trapping potential. The interaction strengths $g_{ij}=4\pi \hbar^2 a_{ij}/m$ and 
$g_{12}=4\pi\hbar^2 a_{12}/m$ are given in terms of the $s$-wave scattering lengths, $a_{ij}$. 
In the following we consider the specific 
case of the two hyperfine levels $|F=1,m_F=\pm 1\rangle$ of $^{23}$Na for which 
$a_{11}=a_{22}=54.54 a_B$ and $a_{12}=50.78 a_B$ (where $a_B$ is the Bohr radius) and we 
define $g=g_{11}=g_{22}$. The last equality means that the system is $\mathbb{Z}_2$ symmetric 
and  the ground state is characterised by $n_1=n_2=n/2$~\footnote{The $\mathbb{Z}_2$ symmetry 
is spontaneously broken if $\delta g$ is enough negative~(see, e.g., \cite{Abad2013})}. A peculiarity 
of sodium atoms is that the two $m_F=\pm 1$ states are miscible ($\delta g=g-g_{12}>0$) and that 
the value $\delta g/g=0.07$ is sufficiently large to allow for the observability of the effects 
discussed in this Letter. These important features satisfied by $^{23}$Na have already proven 
to be important to observe magnetic effects \cite{Gallemi2018,Kang2019} and spin superfluidity 
\cite{Kim2017,Fava2018}.

As already mentioned, the two coupled GPEs (see Supplementary Material, Eq.(D1)) derived by 
minimizing the energy functional (\ref{eq:eq1}) admit, in uniform matter, a stationary solitonic 
solution characterized by a peculiar behavior of the relative phase between the two condensates. 
Indeed under the assumption that the densities are homogeneous and equal, the GPEs can be reduced 
to the simpler sine-Gordon equation $\hbar\nabla^2\phi=m\Omega\sin\phi$ \cite{Son2002} for the 
relative phase $\phi=\phi_1-\phi_2$ between the two condensates. This equation admits the following 
1D solitonic solution \cite{Tabor1989}
\begin{equation}
\phi_{1,2}(x)=\pm2\arctan(\exp(x/\xi_\Omega))\,,
\label{eq:eq2}
\end{equation}
which  connects the values $0$ and $2\pi$ of the relative phase $\phi$ as $x$ goes from 
$-\infty$ to $\infty$. Such a solution generates a counterflow current, accumulated in a small 
region of the relative phase gradient of size $\xi_\Omega =\sqrt{\hbar/m\Omega}$ (Rabi healing 
length), fixed by $\Omega$. The solution (\ref{eq:eq2}) of the sine-Gordon equation is actually 
an approximation to the solution of the coupled GPEs, which still exhibits an equivalent phase 
pattern, but with a density dip near the wall, as a consequence of the compressibility of the 
gas. The dip increases with the increase of the Rabi coupling~\cite{Usui2015}, as shown in Fig. \ref{fig:fig1} (a), and in the case of $^{23}$Na, the central density vanishes for 
$\Omega=20.57\,\Omega_c$, reaching the dark soliton state~\cite{Tsuzuki1971}. 

A major issue concerns the stability of the domain wall. Dark solitons
\cite{Muryshev1999,Pitaevskii2016}, as well as dark-bright solitons \cite{Busch2001}, are 
dynamical unstable, except for one dimensional geometries. As first pointed out by Son and 
Stephanov \cite{Son2002}, the relative phase domain wall is instead dynamically 
stable for $\Omega$ smaller then a critical value $\Omega_c$, even in 3D configurations. 
Under the assumption that $\hbar\Omega,\delta gn \ll gn$, the critical value takes the 
simple expression $\Omega_c=n\delta g /3$~\cite{Son2002}. The (transverse) dynamical stability 
of the domain wall is related to the sign of its effective mass, which fixes the energy change 
of the configuration when the soliton moves with small velocity $v$ according to 
$E_{DW}(v)= E_{DW}^{(v=0)} + m^*v^2/2$, 
where $E_{DW}^{(v=0)}=A^{-1}\int d\vec r\,^3(\mathcal{E}_{\rm GP}[\Psi_{\rm DW}]-\mathcal{E}_{\rm GP}[\Psi_{\rm GS}])$ is the energy cost per unit area ($A$) to create a domain wall at rest. 
Here $\Psi_{\rm GS}$ is the order parameter of a uniform mixture and $\Psi_{\rm DW}$ the one 
in presence of the domain wall. Analytical expression for $m^*$ can be obtained in two limiting 
cases. For $\Omega\ll\Omega_c$, one gets~\cite{Qu2017} $m^*=8m\xi_s n\sqrt{\Omega/\Omega_c}$, 
where $\xi_s=\hbar/\sqrt{2m\delta g n}$  is the spin healing length. In the case 
$\Omega\simeq\Omega_c$ we find that the effective mass per unit area (see Supplementary material, 
Section II) in the incompressible regime ($\hbar\Omega\ll\mu$) can be approximated as
\begin{equation}
m^*=m\frac{\sqrt{3}\pi^2}{4}\xi_s n\frac{\sqrt{\Omega\Omega_c}}{\Omega_c-\Omega} \,,
\label{m*2}
\end{equation}
showing that the effective mass diverges at $\Omega_c$ and becomes negative for 
$\Omega>\Omega_c$.

It was shown in~\cite{Kamchatnov2008} for dark solitons that the low-momentum dispersion 
for the transverse excitations can be written as
\begin{equation}
\lim_{q\to \,0} \frac{\omega(q)}{q}=\sqrt{\frac{E_{\rm DW}^{(v=0)}}{m^*}}\,,
\label{m*3}
\end{equation}
where $q$ is the in-plane wave vector of the excitation. Equation (\ref{m*3}) holds under very 
general assumptions and can be applied also to the domain wall. In particular one sees that for 
$\Omega<\Omega_c$, the positiveness of the effective mass results into a real excitation frequency. 
For larger values of $\Omega$, the effective mass is instead negative, leading to a purely 
imaginary excitation frequency, which indicates the occurrence of a dynamic instability. In 
Fig. \ref{fig:fig1} (c) we report the behavior of the imaginary component of the transverse 
excitations for $\Omega>\Omega_c$ as a function of the dimensionless parameter $q \xi$, with 
$\xi=\hbar/\sqrt{mgn}$, obtained by solving the Bogoliubov equations (see Supplementary material, 
section IV) in the case of $^{23}$Na. The spectrum is reminiscent of the dark soliton 
one~\cite{Muryshev1999}, and converges to it when the central density becomes zero. In 
Fig. \ref{fig:fig1}(d) we also compare 
the value of the effective mass calculated numerically using Eq. (\ref{m*3}) with the prediction of 
the analytic formula (\ref{m*2}). The agreement is very good in a wide range of values of the Rabi 
coupling.

As we will discuss in the second part of the Letter domain walls are most likely generated at a 
finite velocity. In Ref. \cite{Qu2017} it was shown that a moving domain wall is charaterized 
by finite magnetization localized near the wall. Both the density dip, discussed above, and 
the magnetization are important features for the experimental identification of the soliton. 
The direct measurement of the relative phase is in fact of difficult experimental access 
\cite{Matthews1999a}, although the relative flow, i.e., the gradient of $\phi$, could be extracted 
via spin selective spatially resolved Bragg scattering. This technique has been very recently 
applied in Ref. \cite{Kang2019} to identify topological composite objects in a spin-1 Bose gas.

Again in the limit $\hbar\Omega,\delta gn \ll gn$ and for small velocity, an analytical 
expression for the local magnetization $\eta$ can be derived (see Supplementary material, 
Section I):
\begin{equation}
\eta=\frac{n_1-n_2}{n} = -\frac{m^* v}{\hbar n}{\mbox{sech}^2\left(\frac{\Omega}{3\Omega_c\xi_s}(x-vt)\right)},
\label{eta}
\end{equation}
which is shown in Fig. \ref{fig:fig1}(b) in units of $\eta/U$ where $U\equiv v/c_s$ is the velocity in units of the spin speed of sound $c_s=\hbar/2m\xi_s$ calculated for $\Omega=0$.

\begin{figure*}[t!]
\includegraphics[width=\linewidth]{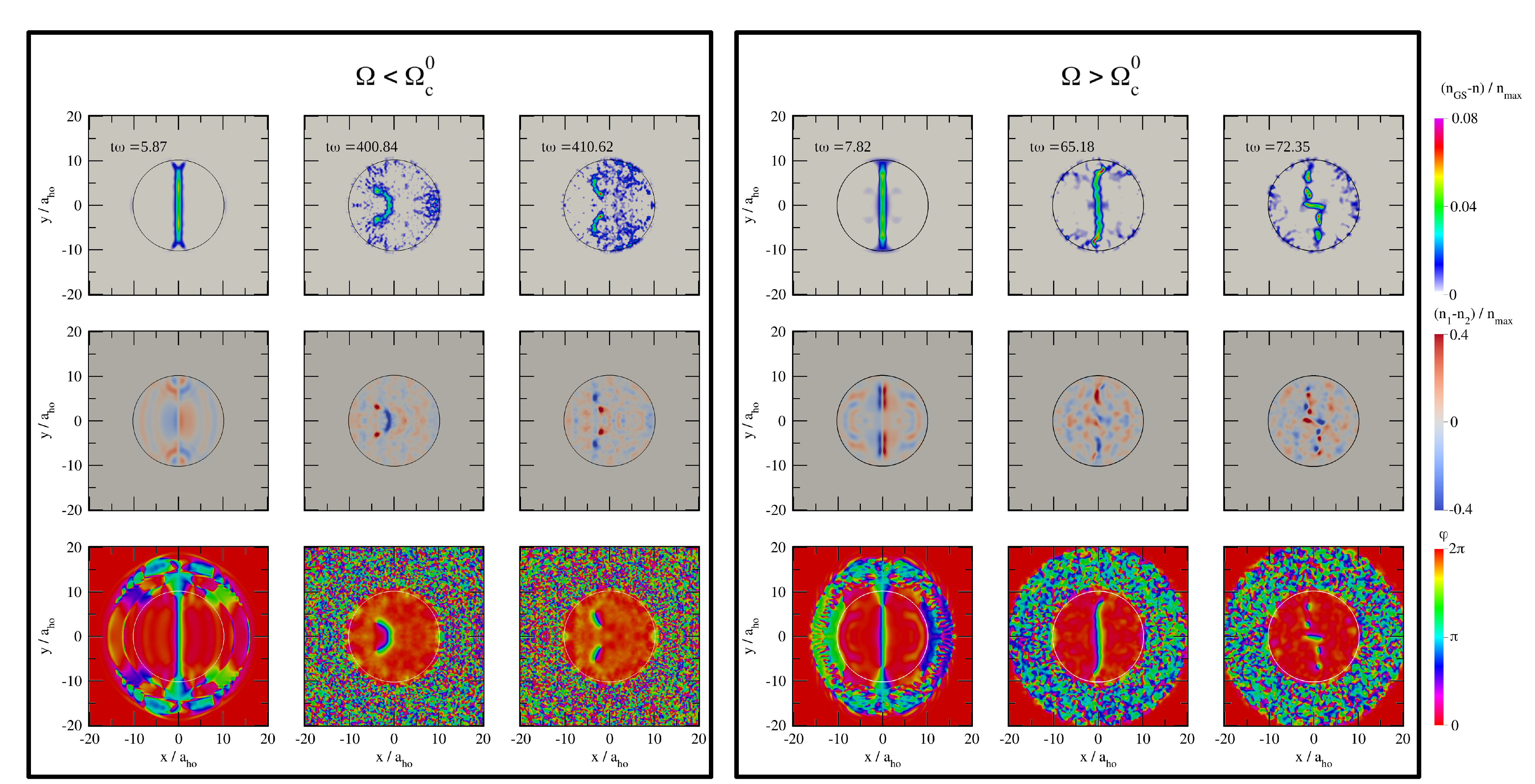}
\caption{The rows show the evolution of the total density with respect to the total density 
of the ground state (first), the magnetization (second) and the relative phase as a color map 
(third), for $\Omega=0.8\,\Omega_c^0$ for times: $t\omega=5.87$ (first column), $t\omega=400.84$  
(second) and $t\omega=410.62$ (third), and $\Omega=1.4\,\Omega_c^0$ for times: $t\omega=7.82$, 
$t\omega=65.18$ and $t\omega=72.35$. The solid (black or white) circles correspond to isocountours 
at $5\%$ of the central ground state density.}
\label{fig:fig2}
\end{figure*}

Having characterized the basic properties of the domain wall in uniform matter we 
are now ready to discuss the possible realization and the study of the different decay mechanisms 
when the mixture is trapped in a harmonic potential. In order to provide a quantitative description 
of the domain wall decay we rely on the solution of time-dependent coupled GPEs. We consider 
$N=8\times10^5$ atoms in a quasi-2D geometry confined by a cylindrically symmetric harmonic 
trap $V_{\rm ext}=m(\omega^2r^2+\omega_z^2z^2)/2$, with $\omega=2\pi\times8\,\mbox{Hz}$ and 
$\omega_z=2\pi\times320\,\mbox{Hz}$ \footnote{In a quasi-2D geometry, the scattering lengths 
are renormalized to $a_i/\sqrt{2\pi l_z^2}$ with $l_z=\sqrt{\hbar/m\omega_z}$}. By evolving 
the GPE in imaginary time, we first prepare the condensate in its ground state in which each atom 
is in a 50/50 superposition of the two hyperfine internal levels. The cloud has the typical inverted parabola density profile and we define for later convenience the central critical Rabi coupling as 
$\Omega_c^0= \delta g n(0)/3$ with $n(0)$ the density in the center of the trap.

A relative phase profile in the form of Eq.~(\ref{eq:eq2}), which crosses the whole cloud at 
the center of the trap, is then imprinted onto the condensate wave functions and let to evolve in 
real time by using the GPEs. Notice that this initial condition should be easily realised in 
current experiments. Since the phase imprinting does not change the total density, the initial 
configuration is not the proper domain wall solution and a local dip is rapidly produced 
(see first and fourth column of Fig. \ref{fig:fig2}). The noise created in this first stage 
of the evolution will play a crucial role in driving the subsequent dynamics of the wall, which 
depends very much on whether the value of $\Omega$ is smaller or larger than $\Omega_c^0$. The 
two cases give rise to very different scenarios for the evolution of the domain wall, both 
concerning the nature of the instability and the corresponding time scales.

When $\Omega<\Omega_c^0$, a local density approach suggests that in the region where
$\Omega<\Omega_c(n(r))$ the domain wall is dynamically stable. For sufficiently small 
values of $\Omega$ the region of dynamical instability is confined to a narrow region near 
the surface (which is shown as a thin line) and we find that a domain wall confining a 
vortex-antivortex pair in the same component is formed after a while. The vortices are 
located symmetrically at a distance $R$ from the center of the trap, near the surface. 
This pair does not carry angular momentum, and starts moving with linear velocity, causing the 
bending of the wall and the emergence of a visible local magnetization (second column of 
Fig. \ref{fig:fig2}). Such a bending  is in agreement with the bending already reported in 
Ref. \cite{Eto2018} in uniform matter. When the bending becomes pronounced the domain wall 
breaks down into a certain number of vortex-vortex and antivortex-antivortex pairs and, 
depending on the actual value of $\Omega$, an additional vortex-antivortex pair, that rapidly 
moves towards the cloud boundary, where gets absorbed or breaks into other pairs. A typical 
frame of such rich dynamics is shown in the third column of Fig. \ref{fig:fig2}.

When $\Omega>\Omega_c^0$ a vortex-vortex pair appears, which starts rotating, dragging the edges 
of the wall (see Fig.~\ref{fig:fig2}, fifth column). However, the domain wall is dynamically 
unstable similarly to the homogeneous case and the noise caused by the phase imprinting procedure 
triggers the decay, leading to the snaking of the domain wall. The simulation shows that on short 
times this snaking nucleates multiple domain walls of smaller size which confine pairs of vortices 
of equal vorticity, belonging to different atomic species (sixth column of Fig. \ref{fig:fig2}). 
These smaller objects are more stable since the minimum length required to exhibit dynamic 
instability is given by $\pi /q_0$, where $q_0$ is the threshold value of the wave vector 
below which the Bogoliubov excitation frequency becomes imaginary (see Fig. \ref{fig:fig1}(c) 
and Supplementary Material, Sections III and IV).

As reported in Fig. \ref{fig:fig2}, it is evident that the domain wall breaking for 
$\Omega<\Omega_c^0$ is almost an order of magnitude slower than for $\Omega>\Omega_c^0$, 
reflecting the energetic nature of the instability. With the chosen values of the trap 
fequencies the decay time varies from  1 to 10 seconds in the two cases reported in 
Fig. \ref{fig:fig2}. Even assuming that such time scale is reduced by one or two orders 
of magnitude in a real experiment, such a phenomenon should be still measurable. 

For completeness in the first two rows of Fig. \ref{fig:fig2}, we also show the behavior 
of the total density and of the magnetization for the same temporal sequence. These 
quantities are measurable experimentally (typical values of the dip reported in Fig. \ref{fig:fig2} 
correspond to a contrast of about 5\% in the central region) and can therefore directly 
probe the nature of the decay and of the corresponding dynamical evolution of the vortex 
pairs, given the strong correlation between the density dip, the magnetized wall arising at 
$\Omega<\Omega_c^0$ and the kink in the relative phase.

The initial conditions discussed so far, based on the phase imprinting of the domain wall is 
not the only possibility that one can explore theoretically. In order to check the general 
validity of the instability mechanisms, we have considered two additional routes 
for generating the decay. In the first case we initially create an infinitely 
long domain wall corresponding to the exact stationary solution of the GPEs. Being exact, 
such a metastable configuration would not evolve in time unless a small amount of noise is 
added to the system. We have analyzed this possibility by evolving in time the wave function 
$\beta(\vec r\,)\Psi_{\rm DW}(\vec r\,)$, where $\Psi_{\rm DW}(\vec r\,)$ is the exact domain 
wall solution and $\beta(\vec r)$ is a random number in the range $(1-\varepsilon,1+\varepsilon)$, 
$\varepsilon$  being the noise strength ($1\gg\varepsilon>0$). The presence of additional 
noise triggers the dynamical instability for $\Omega>\Omega_{c}^0$ and it is indispensable for 
$\Omega<\Omega_c^0$ in order to favour the transition to lower energy configurations characterized 
by a domain wall of smaller size connecting a vortex pair, which is protected by an energy 
barrier due to the presence of the trap (see Supplementary Material, Section V). Alternatively 
instead of adding noise to the initial configuration, we produce a fast ramping of $\Omega$, 
which sets the system out of equilibrium. Such a protocol would allow to improve the visibility 
of the magnetized strings and the density dip. In both cases, very similar dynamics take place.

In conclusion, we have investigated the decay mechanisms exhibited by a relative phase domain 
wall in a coherently-coupled condensate, with special emphasis to the role played 
by the external harmonic trapping. When the Rabi frequency is larger than the critical value 
above which domain walls are dynamically unstable, the snake instability mechanism triggers 
the decay, leading to the breaking of the domain wall into confined vortex pairs. In 
contrast, when the Rabi frequency is below the critical one the instability of the soliton has 
an energetic nature and additional excitations are required to allow the entrance of the vortices 
from the border of the trapped atomic cloud. The novel scenarios discussed in this Letter are 
expected to open challenging perspectives to be explored in future experiments on coherently 
coupled atomic gases to obtain composite topological objects \cite{Kang2019} and mimic pair 
creation and string breaking.

\paragraph*{Note added:}During the final writing of this Letter we have become aware of a 
new theoretical work which addresses the problem of the dynamic instability of the relative 
phase domain walls in uniform matter \cite{Ihara2019}, in general agreement with some of 
our findings. 

\begin{acknowledgments}

This project has received funding from the European Union's Horizon 2020 research and innovation 
programme under grant agreement No. 641122 ``QUIC'', the FIS$\hbar$ project of the Istituto 
Nazionale di Fisica Nucleare and the Provincia Autonoma di Trento. We would like to thank 
useful discussions with Gabriele Ferrari and Giacomo Lamporesi. 
\end{acknowledgments}

\newpage

\appendix

In the following Supplementary Material we derive some useful results for
the magnetization, the effective mass and the frequency of the elementary
excitations associated with the domain wall near the critical point $\Omega_c$, 
which have been discussed in the main text. In Section \ref{app:barrier} we also 
derive explicit results for the barrier encountered by a vortex pair before 
entering a harmonically trapped atomic cloud.

\section{Velocity-induced magnetization}
\label{sect:magn}
In this section we calculate the magnetization of a slowly moving domain 
wall. Let us define $\kappa^{2}=\Omega /3\Omega_{c}=\xi_{s}/\xi_{\Omega}$, where 
$\xi_{s}=\hbar/\sqrt{2m\delta gn}$ and $\xi_{\Omega}=\sqrt{\hbar/m\Omega}$ are the spin 
and the Rabi healing lengths respectively (see main text) and introduce $U=v/c_{s}$ 
the velocity of the domain wall in units of the spin sound speed $c_{s}=\sqrt{\delta gn/2m}$, 
and $n\sin \eta =n_{1}-n_{2}$ the magnetization of the atomic cloud. In the following 
analysis we always ignore the density dip exhibited by the domain wall and make us of the 
Eqs. (15) and (16) of Ref. \cite{Qu2017}: 
\begin{equation}
U\frac{\partial \eta }{\partial \zeta }-2\sin \eta \frac{\partial \eta }{%
\partial \zeta }\frac{\partial \phi _{R}}{\partial \zeta }+\cos \eta \frac{%
\partial ^{2}\phi _{R}}{\partial \zeta ^{2}}=\kappa ^{2}\sin \phi _{R}\,,
\label{appeq:b1}
\end{equation}%
and 
\begin{align}
U\cos \eta \frac{\partial \phi _{R}}{\partial \zeta }=& \frac{\partial
^{2}\eta }{\partial \zeta ^{2}}-\sin \eta \cos \eta \left( 1-\left( \frac{%
\partial \phi _{R}}{\partial \zeta }\right) ^{2}\right)  \notag \\
& -\kappa ^{2}\sin \eta \cos \phi _{R}\,,  
\label{appeq:b2}
\end{align}%
where $\zeta =(x-vt)/\xi _{s}$ and $\phi _{R}$ is the relative phase of the
domain wall. We will solve the above equations
considering small deviations from the Son-Stephanov stationary solution 
\begin{equation}
\phi =\phi _{1}-\phi _{2}=4\arctan (\exp (\kappa \zeta ))\,.  
\label{A3}
\end{equation}%
The perturbative calculation requires $\phi _{R}-\phi =\chi \ll 1$ and $\eta
\ll 1$. It is easy to show that to lowest order, $\eta \sim U$ and $\chi
\sim U^{2}$. Thus, one can simplify Eq. (\ref{appeq:b2}) to the form
\begin{equation}
-\frac{\partial ^{2}\eta }{\partial \zeta ^{2}}-\eta \left[ \left( \frac{%
\partial \phi }{\partial \zeta }\right) ^{2}-1-\kappa ^{2}\cos \phi \right]
=-U\frac{\partial \phi }{\partial \zeta }\,,  
\label{appeq:b3}
\end{equation}%
where $\partial _{\zeta }\phi =2\kappa /\cosh (\kappa \zeta )$. Using the identity 
\begin{equation}
\kappa ^{2}\cos \phi =\kappa ^{2}+\left( \partial _{\zeta }\phi
\right) ^{2}-\frac{6\kappa ^{2}}{\cosh ^{2}(\kappa \zeta )}\,
\label{iden}
\end{equation}
we can rewrite Eq. (\ref{appeq:b3}) as: 
\begin{equation}
\hat{\mathcal{O}}\eta =-U\frac{\partial \phi }{\partial \zeta }\,,
\label{O1}
\end{equation}
where the P\"{o}schl-Teller-like operator 
\begin{equation}
\hat{\mathcal{O}}=-\frac{\partial ^{2}}{\partial \zeta ^{2}}-\frac{6\kappa
^{2}}{\cosh ^{2}(\kappa \zeta )}+1+\kappa ^{2}  
\label{PTL}
\end{equation}
has the lowest eigenvalue $1-3\kappa ^{2}$ with its normalized eigenvector 
\begin{equation}
\psi _{\mathcal{O}}(\zeta )=\frac{\sqrt{3\kappa }}{2\cosh ^{2}(\kappa \zeta )}\,,  
\label{A7}
\end{equation}%
corresponding to the bound solution of even parity. The exact solution of Eq.~(\ref{O1}) can be 
presented in the form 
\begin{equation}
\eta (\zeta )=\mathcal{C}\psi _{\mathcal{O}}(\zeta )+\delta \eta \,,
\label{deltaeta}
\end{equation}
where $\delta \eta $ is the contribution of the delocalized states, orthogonal to 
$\psi _{\mathcal{O}}$ . By multiplying both members of Eq. (\ref{O1}) by 
$\psi _{\mathcal{O}}^* =$ $\psi _{\mathcal{O}}$, and integrating over $\zeta $ from 
$-\infty $ to $\infty $, the proportionality constant $\mathcal{C}$ is found to be 
\begin{equation}
\mathcal{C}=-\frac{U}{1-3\kappa ^{2}}\int_{-\infty }^{\infty }d\zeta \,\psi
_{\mathcal{O}}^*(\zeta )\,\frac{\partial \phi }{\partial \zeta }=-%
\frac{\sqrt{3}\,\pi }{2}\frac{\sqrt{\kappa }\,U}{1-3\kappa^2}\,.  
\label{appeq:b8}
\end{equation}%
This expression diverges for $\kappa =1/\sqrt{3}$ (the critical point), which suggests that 
the contribution to the magnetization (\ref{deltaeta}) that comes from the delocalized states 
can be indeed neglected. Notice that in order to fulfill the condition $\eta \ll 1$ the velocity 
of the soliton should satisfy the condition ${U\ll 1-3\kappa ^{2}}$.

\section{Effective mass of a domain wall}

In order to calculate the effective mass we start from the general expression of the energy of 
the domain wall per unit area, given in Ref. \cite{Qu2017}: 
\begin{align}
E& =\frac{n\hbar c_{s}}{4}\int d\zeta \left[ \left( \frac{\partial \eta }{%
\partial \zeta }\right) ^{2}+\cos ^{2}\eta \left( \frac{\partial \phi _{R}}{%
\partial \zeta }\right) ^{2}\right.   \notag \\
& +\sin ^{2}\eta +2\kappa ^{2}(1-\cos \eta \cos \phi _{R})\bigg]\;,
\end{align}%
where we have again neglected the presence of the density dip. Keeping only
terms of order $U^{2}$ (that is $\eta ^{2}$ or $\chi$, the terms linear 
in $\chi $ being vanishing), we can obtain the second order term in the 
expansion of the energy as a function of the velocity, hereafter called $E^{(2)}$: 
\begin{equation}
E^{(2)}=\frac{n\hbar c_{s}}{4}\int d\zeta \left[ \left( \frac{\partial \eta 
}{\partial \zeta }\right) ^{2}-\eta ^{2}\left( \frac{\partial \phi }{%
\partial \zeta }\right) ^{2}+\eta ^{2}+\kappa ^{2}\eta ^{2}\cos \phi \right]
\,.  \label{appeq:c2}
\end{equation}%
By inserting in Eq. (\ref{appeq:c2}) the Son-Stephanov expression (\ref{A3}) for the relative 
phase $\phi $ of the moving domain wall and performing partial integration, one can rewrite 
(\ref{appeq:c2}) as 
\begin{align}
E^{(2)}=& \frac{n\hbar c_{s}}{4}\int d\zeta \,\eta \,\hat{\mathcal{O}}\,\eta
=\frac{n\hbar c_{s}}{4}(1-3\kappa ^{2})\,\mathcal{C}^{2}=  \notag \\
& \frac{3\pi ^{2}}{16}\frac{n\hbar \kappa v^{2}}{c_{s}}\frac{1}{1-3\kappa
^{2}}\,,
\end{align}%
where we have used the ansatz (\ref{deltaeta}) with $\delta\eta=0$ and the approximate 
result (\ref{appeq:b8}) for the constant $\mathcal{C}$.

Thus, the effective mass per unit area, defined by:
\begin{equation}
m^*=\lim_{v\to 0}\frac{1}{v}\frac{dE(v)}{dv}\,,
\end{equation}
takes the simple insightful form: 
\begin{equation}
m^*=m\frac{3\pi^2}{4}\frac{\xi_s n \kappa}{1-3\kappa^2}\,.  
\label{appeq:c5}
\end{equation}
This is the line shown as a red dashed line in Fig. 1(d) in the main text. The limit of validity of 
this equation is exactly the same as in the calculation of the magnetization (see previous section), 
being based on the approximated solution (\ref{A7}), yielding result (\ref{appeq:b8}) for the relevant 
constant $\mathcal{C}$.

\section{Imaginary frequencies of the excitation spectrum}

In order to get a good approximation for the full $q$-dependence of the imaginary frequency: 
\begin{equation}
\omega(q) = i \gamma (q)
\end{equation}
that arises at $\Omega>\Omega_c$, one has to consider not only the fluctuations of the magnetization 
and the relative phase, but also the fluctuations of the total phase $\phi_T=\phi_1+\phi_2$, which 
are particularly relevant at large wave vectors.

To this aim, we will start from the Lagrangian previously introduced in Ref. \cite{Qu2017}, 
including not only the magnetization and the relative phase but also the total phase: 
\begin{align}
\mathcal{L}=& \sin \eta \frac{\partial \phi _{R}}{\partial \tau }-\frac{g}{\delta g}+
\frac{1}{2}\cos ^{2}\eta  \notag \\
& -\frac{1}{2}\left[ (\nabla \phi _{R})^{2}+(\nabla \phi _{T})^{2}+(\nabla
\eta )^{2}-2\sin \eta \nabla \phi _{R}\nabla \phi _{T}\right]  \notag \\
& +\kappa ^{2}\cos \eta \cos \phi _{R}\,.
\end{align}%
One can then obtain the corresponding equations of motion for each of the
variables $\mathcal{F}=\{\eta ,\phi _{R},\phi _{T}\}$, using the Lagrange
equation: 
\begin{equation}
\frac{\partial \mathcal{L}}{\partial \mathcal{F}}-\nabla \frac{\partial 
\mathcal{L}}{\partial (\nabla \mathcal{F})}-\frac{d}{d\tau }\frac{\partial 
\mathcal{L}}{\partial (\partial _{\tau }\mathcal{F})}=0\,,
\end{equation}%
from which one can obtain the following set of the equations: 
\begin{align}
& \frac{\partial \eta }{\partial \tau }+\frac{\partial ^{2}\chi }{\partial
\zeta ^{2}}-q^{2}\xi_s^2\chi -\nabla (\eta \nabla \phi _{T})=\kappa ^{2}\cos \phi
_{R}  \notag \\
& \frac{\partial \chi }{\partial \tau }+\frac{\partial ^{2}\eta }{\partial
\zeta ^{2}}-q^{2}\xi_s^2\eta +\nabla \chi \nabla \phi _{T}+\nabla \phi _{R}\nabla
\beta -\eta =\kappa ^{2}\eta \cos \phi _{R}  \notag \\
& \frac{\partial ^{2}\beta }{\partial \zeta ^{2}}-q^{2}\xi_s^2\beta -\nabla(\eta \nabla\phi _{R})=0\,,
\end{align}%
where we have expressed the fluctuations as: 
\begin{align}
\eta (\vec{r}\,,\tau )=& \eta (\zeta )\exp (ir_{\perp }q\xi_s+\gamma \tau ) 
\notag \\
\phi _{R}(\vec{r}\,,\tau )=& \phi (\zeta )+\chi (\zeta )\exp (ir_{\perp
}q\xi_s+\gamma \tau )  \notag \\
\phi _{T}(\vec{r}\,,\tau )=& \beta (\zeta )\exp (ir_{\perp }q\xi_s+\gamma \tau
)\,.
\end{align}%
By working to order $U$ (or equivalently $\eta $), the previous equations
take the simplified form:  
\begin{align}
& \gamma \eta +\frac{\partial ^{2}\chi }{\partial \zeta ^{2}}-q^{2}\xi_s^2\chi
-\kappa ^{2}\cos \phi =0  
\label{eq:app:d5} \\
& \gamma \chi +\frac{\partial ^{2}\eta }{\partial \zeta ^{2}}-q^{2}\xi_s^2\eta +%
\frac{\partial \phi }{\partial \zeta }\,\frac{\partial \beta }{\partial
\zeta }-\eta -\kappa ^{2}\eta \cos \phi =0  
\label{eqeta} \\
& \frac{\partial ^{2}\beta }{\partial \zeta ^{2}}-q^{2}\xi_s^2\beta -\frac{\partial 
}{\partial \zeta }(\eta \,\frac{\partial }{\partial \zeta }\phi )=0\,.
\label{eqbeta}
\end{align}%
Let us now solve the first equation, by writing $\chi (\zeta )=\mathcal{D}%
\psi _{\mathcal{P}}(\zeta )+\delta \chi $, where $\mathcal{D}$ is a proportionality coefficient 
(the final result will not depend on its value) and 
\begin{equation}
\psi _{\mathcal{P}}(\zeta )=-\frac{\sqrt{\kappa /2}}{\cosh (\kappa \zeta )}
\label{zeromode}
\end{equation}%
is the only bound eigenvector of the P\"{o}schl-Teller operator 
\begin{equation}
\hat{\mathcal{P}}=-\frac{\partial ^{2}}{\partial \zeta ^{2}}-\frac{2\kappa
^{2}}{\cosh ^{2}(\kappa \zeta )}+\kappa ^{2}
\end{equation}%
whose eigenvalue is $0$. Notice that the zero-mode solution (\ref{zeromode}) has a 
simple physical meaning. This is a change of the Son-Stephanov static solution 
(\ref{A3}) after an infinitesimal small displacement along $\zeta$ direction.

The function $\delta \chi $ is a combination of the other eigenstates that are, by construction, 
orthogonal to $\psi _{\mathcal{P}}$. Analogously to what we did in Section \ref{sect:magn} for the 
magnetization, we can multiply both members of Eq. (\ref{eq:app:d5}) by $\psi _{\mathcal{P}}^*$ 
and integrate over all space to get: 
\begin{equation}
\gamma M\mathcal{C}=\mathcal{D}q^{2}\xi_s^2\,,  
\label{appeq:eq11}
\end{equation}
where 
\begin{equation}
M\equiv \int_{-\infty }^{\infty }d\zeta \,\psi _{\mathcal{P}}^{\ast }\,\psi
_{\mathcal{O}}=\frac{\sqrt{3}}{4\sqrt{2}}\pi \,.
\end{equation}

\begin{figure}[t!]
\includegraphics[width=\linewidth]{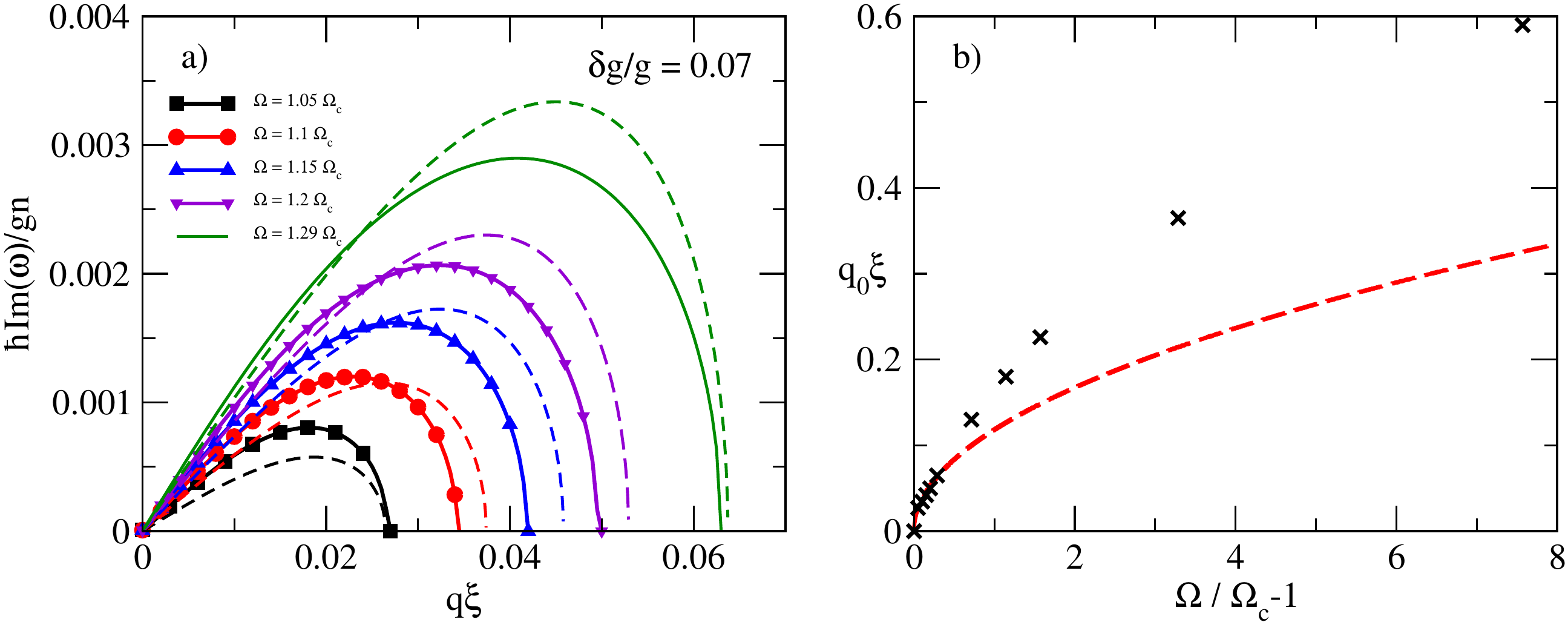}
\caption{a) Imaginary frequencies for different values of the Rabi frequency
calculated numerically (solid lines with symbols) and analytically by Eq. 
(\ref{eq:app:d19}) (dashed lines). b) Value of the finite momentum at
which the imaginary frequency vanishes $q_0$, as a function of $%
\Omega/\Omega_c$, calculated numerically (crosses) and analytically (red
dashed line). }
\label{appfig:fig1}
\end{figure}

The other two equations are coupled. In order to decouple them, we first
transform Eq. (\ref{eqeta}) with help of Eq. (\ref{iden}). One gets:
\begin{equation}
\gamma \chi -\hat{\mathcal{O}}\eta -q^{2}\xi_s^2\eta -\eta \left( \frac{\partial
\phi }{\partial \zeta }\right) ^{2}+\frac{\partial \phi }{\partial \zeta }\,%
\frac{\partial \beta }{\partial \zeta }=0  \label{eqeta1}
\end{equation}%
Now let us multiply this equation on $\psi _{\mathcal{O}}^{*}$ and
integrate over the space. The result is 
\begin{align}
M\mathcal{D}\gamma  -\mathcal{C}(3\kappa ^{2}-1-q^{2}\xi_s^2)=  \nonumber \\
\mathcal{C}\int_{-\infty }^{\infty }
d\zeta \left\vert \psi _{\mathcal{O}}\right\vert ^{2}\left( \frac{\partial
\phi }{\partial \zeta }\right) ^{2}-\int_{-\infty }^{\infty } d\zeta \psi _{\mathcal{O}}^{\ast }\left( \frac{\partial
\phi }{\partial \zeta }\frac{\partial \beta }{\partial \zeta }\right) 
\label{eqC}
\end{align}
where we have ignored smaller $\delta \chi $ and $\delta \eta $ terms. This equation still depends on 
$\beta$, but it can be further simplified if we take into account that because of the fast 
decreasing of the functions $\psi _{\mathcal{O}}$ and $\partial_\zeta\phi$ as $\vert\zeta\vert\to\infty$, 
the integrals of the right hand side of Eq. (\ref{eqC}), are relevant only when 
$\zeta\lesssim \kappa^{-1}$. On the other hand, it is easy to show that the term proportional 
to $q^{2}$ in Eq. (\ref{eqbeta}) is important only on distances larger than $q^{-1}\xi_s^{-1}$. 
However, near the critical point the instability takes place only for 
$q<q_{0}\propto \sqrt{3\kappa^2-1}\ll 1$ (see below). This means that in the integration (\ref{eqC}) 
one can calculate $\beta$ neglecting the term proportional to $q^{2}$ in Eq. (\ref{eqbeta}), 
which leads to:
\begin{equation}
\frac{\partial ^{2}\beta }{\partial \zeta ^{2}}-\frac{\partial }{\partial
\zeta }(\eta \,\frac{\partial }{\partial \zeta }\phi )=0\,.
\label{eqD}
\end{equation}
Integration of Eq. (\ref{eqD}) yields
\begin{equation}
\frac{\partial \beta }{\partial \zeta }=\eta \,\frac{\partial \phi }{%
\partial \zeta }\approx \mathcal{C}\psi _{\mathcal{O}}\frac{\partial \phi }{\partial
\zeta }\,,
\end{equation}%
and substitution into Eq. (\ref{eqC}) yields to the cancellation of the two terms in the right hand side. Finally, 
we obtain the simple equation:
\begin{equation}
M\mathcal{D}\gamma -\mathcal{C}(3\kappa ^{2}-1-q^{2}\xi_s^2)=0\,.
\label{eqC2}
\end{equation}

In conclusion, Eq. (\ref{eqC2}) can be combined with Eq. (\ref{appeq:eq11}) to give the 
final result: 
\begin{equation}
\gamma =\frac{4\sqrt{2}}{\sqrt{3}\pi }(3\kappa ^{2}-1-q^{2}\xi_s^2)^{1/2}q\xi_s\,  
\label{eq:app:d19}
\end{equation}%
for the imaginary part of the frequency holding for $\Omega >\Omega _{c}$.

Figure \ref{appfig:fig1} (a) shows the imaginary part of the
eigenfrequencies for $\Omega$ larger but close to $\Omega_c$, as a function
of the wave vector $q$, for different values of the Rabi frequency. The
solid lines with symbols correspond to the results obtained from the
numerical solution of the Bogoliubov equations (see section \ref%
{app:bogoliubov}), and the dashed lines with the same colors correspond to
the curves given by Eq. (\ref{eq:app:d19}). Panel (b) shows the value of $%
q_0 $, which is the threshold point at which the imaginary frequencies vanish at
finite $q$, as a function of $\Omega/\Omega_c$. Crosses correspond to
numerics and the red dashed line corresponds to the analytical prediction
given by Eq. (\ref{eq:app:d19}).

\section{Bogoliubov equations}

\label{app:bogoliubov}

From the energy functional (Eq. (1) in the main text), one can obtain the corresponding coupled 
Gross-Pitaevskii equations for the two wave functions $\psi_i$ as: $i\hbar\partial_t\psi_i=\delta\mathcal{E}/\delta\psi_i^*$. Static solutions (like the domain wall), obey 
$\psi_i=\psi_i\exp(i\mu t/\hbar)$, and thus can be computed by solving: 
\begin{align}
\mu\psi_1=&-\frac{\hbar^2}{2m}\nabla^2\psi_1+V_{\mathrm{ext}%
}\psi_1+g_{11}|\psi_1|^2\psi_1  \notag \\
&+g_{12}|\psi_2|^2\psi_1-\frac{1}{2}\hbar\Omega\psi_2  \notag \\
\mu\psi_2=&-\frac{\hbar^2}{2m}\nabla^2\psi_2+V_{\mathrm{ext}%
}\psi_2+g_{22}|\psi_2|^2\psi_2  \notag \\
&+g_{12}|\psi_1|^2\psi_2-\frac{1}{2}\hbar\Omega\psi_1\,.  
\label{GPErabi}
\end{align}
Solutions of the previous equations can be directly seen in Figs. 1(c) in the main text and 
\ref{appfig:fig1}(a) in the Supplementary Material in the absence of the external trapping 
($V_{\mathrm{ext}}=0$).

From the Gross-Pitaevskii equations, one derives the Bogoliubov equations by
analyzing the fluctuations on the following static solution $\psi_i\to\psi_i+u_i\exp(i%
\omega t/\hbar)+v^*_i\exp(-i\omega t/\hbar)$, where $u_i(\vec r\,)$ and $%
v_i(\vec r\,)$ correspond to the amplitudes of the fluctuations. For the
domain wall solution, both amplitudes can be factorized by symmetry reasons,
as $u_i(\vec r\,)=\exp(iq\rho)\,u_i(x)$ and $v_i(\vec
r\,)=\exp(iq\rho)\,v_i(x)$, respectively. By introducing the equilibrium
state $\psi(x)$, numerically obtained by solving the Gross-Pitaevskii
equation, the Bogoliubov equations 
\begin{equation}
\mathcal{L} 
\begin{pmatrix}
u_1 \\ 
u_2 \\ 
v_1 \\ 
v_2%
\end{pmatrix}
=\hbar\omega 
\begin{pmatrix}
u_1 \\ 
u_2 \\ 
v_1 \\ 
v_2%
\end{pmatrix}
\,.
\end{equation}
can be solved by diagonalizing the matrix 
\begin{widetext}
\begin{equation}
\mathcal{L}=
\begin{pmatrix}
    B_1 & -\frac{1}{2}\hbar\Omega+g_{12}\psi_1\psi^*_2 & g\psi^2_1 & g_{12}\psi_1\psi_2 \\
   -\frac{1}{2}\hbar\Omega+g_{12}\psi^*_1\psi_2 & B_2 &  g_{12}\psi_1\psi_2 & g\psi^2_2 \\
   -g(\psi^*_1)^2 & -g_{12}\psi^*_1\psi^*_2 & -B_1 & \frac{1}{2}\hbar\Omega-g_{12}\psi^*_1\psi_2 \\
   -g_{12}\psi^*_1\psi^*_2 & -g(\psi^*_2)^2 & \frac{1}{2}\hbar\Omega-g_{12}\psi_1\psi^*_2 & -B_2
\end{pmatrix}
\,.
\end{equation}
\end{widetext}
with the operators $B_i$ defined by : 
\begin{equation}
B_i=\frac{\hbar^2}{2m}\bigg(q^2-\frac{d^2}{dx^2}\bigg)+2g|\psi_i|^2+g_{12}|%
\psi_{3-i}|^2-\mu\,
\end{equation}
and the chemical potential given by 
\begin{equation}
\mu=\frac{1}{2}[(g+g_{12})n-\Omega]\,.
\end{equation}

Diagonalization of this matrix allows us to obtain the eigenfrequencies $%
\omega$, which turn out to be real for $\Omega<\Omega_c$ for any value of $q$
and imaginary for $\Omega>\Omega_c$ and $q$ below $q_0$, which depends on $%
\Omega$. The dependence of the $\mbox{Im}(\omega)$ as a function of $q$ is
very similar to that appearing for the dark soliton, with a slope at $q\to 0$
increasing with $\Omega$. As a consequence, domain walls are dynamically
unstable for $\Omega>\Omega_c$.

\begin{figure}[h!]
\centering
\includegraphics[width=0.7\linewidth]{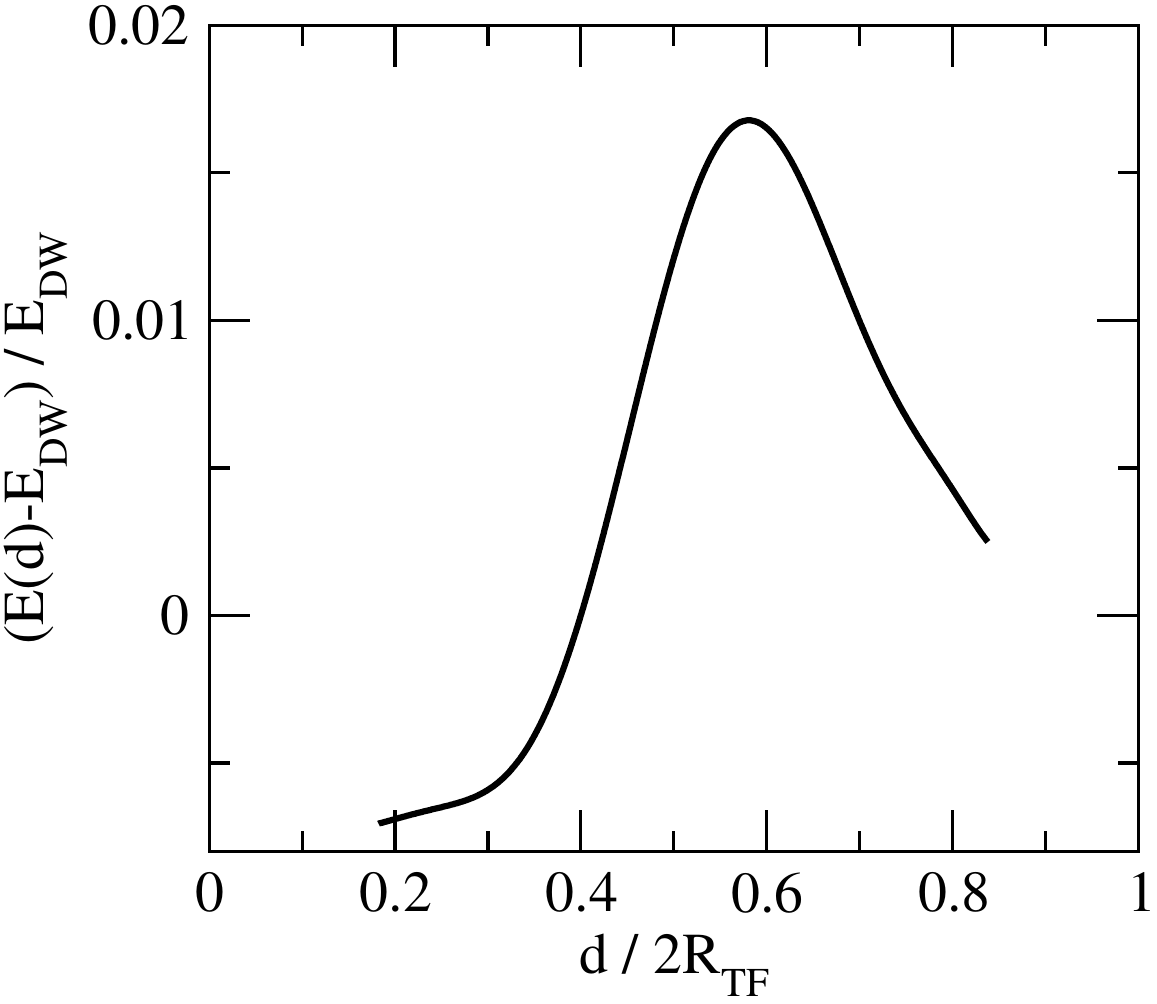}
\caption{Energy of a domain wall binding two vortices separated a distance $%
d $ in a condensate in a harmonic trap with Thomas-Fermi radius $R_{\mathrm{%
TF}}$.}
\label{appfig:fig2}
\end{figure}

\section{Energy of a domain wall as a function of its length in a trap}

\label{app:barrier}

In this section we study the dependence of the energy of a domain wall as a
function of its length localized along one of the diametral axis of an
isotropic harmonic trap. The authors of Ref. \cite{Eto2018} have found that
in uniform matter, the critical length above which the domain wall is
unstable, is much longer for the vortex-vortex than for the
vortex-antivortex configuration. Under the presence of the harmonic trap, a
new effect appears, playing a crucial role in the decay of the domain wall.

The space-dependence of the density in fact gives a strong modification of
the energy as a function of the length, which is specially important when 
the gradient of the density is sharp, i.e. near the boundary. In practice 
the new effect results in an energy barrier that prevents the vortices that 
live at the edge of the condensate to enter inside the central region of 
higher density, unless a thermal or quantum tunnelling mechanism allows for 
such an entrance.

In Fig. \ref{appfig:fig2} we show the dependence of the energy as a function
of the length of a domain wall connecting two vortices in a vortex-vortex
configuration in a gas of $8\times10^5$ $^{23}$Na atoms confined in a
harmonic trap with a trapping frequency equal to $\omega=2\pi\times8\,%
\mbox{Hz}$, and $\Omega=1.2\,\Omega_c$. The figure clearly shows the
existence of an energy barrier when the length of the domain wall approaches
the size of the condensate. It can be understood in terms of the energy cost
of introducing the core of a vortex into the higher-density region. A
similar picture has been also calculated for the vortex-antivortex
configuration, for which we find essentially identical results.

\bibliography{Bibtex.bib}

\end{document}